\documentclass[12pt,preprint]{aastex}

\usepackage{color}
\usepackage{psfig}
\shorttitle{Non-gaussian statistics and stellar systems}
\shortauthors{J.C. Carvalho et al}

\begin{document}

\title{Non-gaussian statistics and stellar
rotational velocities of main sequence field stars}

\author{J.~C.~Carvalho$^{1}$, J.~D.~do Nascimento Jr$^{1}$, R.~Silva$^{1,2}$and J.~R.~De~Medeiros$^{1}$}
\affil{ $^{1}$Universidade Federal do Rio Grande do Norte, UFRN, Departamento
de F\'{\i}sica C. P. 1641, Natal - RN, 59072-970, Brazil}
\affil{ $^{2}$Universidade do Estado do Rio Grande do Norte, 59610-210, Mossor\'o, RN, Brasil}

%% Mark off your abstract in the ``abstract'' environment. In the manuscript
%% style, abstract will output a Received/Accepted line after the
%% title and affiliation information. No date will appear since the author
%% does not have this information. The dates will be filled in by the
%% editorial office after submission.

\begin{abstract}
In this letter we study the observed distributions of rotational
velocity in a sample of more than 16,000 nearby F and G dwarf
stars, magnitude complete and presenting high precision $Vsin i$
measurements. We show that the velocity distributions cannot be
fitted by a maxwellian. On the other hand, an analysis based on
both Tsallis and Kaniadakis power-law statistics is by far the
most appropriate statistics and give a very good fit. It is also
shown that single and binary stars have similar rotational
distributions. This is the first time, to our knowledge, that
these two new statistics are tested for the rotation of such a
large sample of stars, pointing solidly to a solution of the
puzzling problem on the function governing the distribution of
stellar rotational velocity
\end{abstract}


\section{Introduction}

Rotation is a fundamental observable for the study of evolution of
stars, providing also valuable informations on stellar magnetism,
mixing of chemical abundance in stellar interior and tidal
interaction in close binaries. In addition, it can be argued that
the value of the rotational velocity of a star at a given
evolutionary stage may be a measure of its original angular
momentum. In that case, the distribution of rotational velocity
may also be used to study some of the characteristics of the
physical processes controlling star formation. In spite of a large
observational effort carried out by different authors along the
past 50 years, the problem concerning the nature of the
statistical law controlling the distribution of stellar rotational
velocity has turned into a very puzzling question. On the other
hand, there is widespread recognition that stellar rotation axes
have a random orientation (Struve 1945).

The first to derive analytically the distribution of stellar
projected rotational velocity, on the basis of a Gaussian
distribution were Chandrasekhar and Munch (1950). These authors
first assumed a parametric form for a function $f(v)$, where $v$
is the true rotational velocity, then computed the corresponding
distribution of the projected rotational velocity $V sin i$ and
finally adjusted a set of stellar parameters to reproduce the $V
sin i$ measurements. Although Deutsch (1970) has claimed that the
distribution of stellar rotational velocities follows a
Maxwellian-Boltzmann law, a number of studies have shown a clear
discrepancy between theory and observations, where observed
distributions are not well fitted by a Gaussian or Maxwellian
function (e.g.: Wolff et al. 1982; de Medeiros et al. 1996). More
recently, Soares et al. (2006) have shown, based on an analysis of
the rotation of low-mass stars in the Pleiades open cluster, that
the question of the nature of the distribution of stellar
rotational velocity is not simply a question of which mathematical
function model is used. It depends on the statistical mechanics
applied.

Here, we investigate the effects of powerlaw
statistics on the observed distribution of projected rotational
velocity measurements of a sample of more than 16,000 nearby F and
G dwarf stars from the Nordstr$\ddot{o}$m et al. (2004) catalogue.
We show clearly that the velocity distributions cannot be fitted by a
maxwellian. Our analysis is based on both Tsallis and Kaniadakis
non-gaussian statistics and it is shown that these are by far the
most appropriate statistics, giving a very good fit. We also show
that single and binary stars have essentially similar rotational
distributions.

This Letter is organized as follows. In Sec II, based on the
formalism presented by Deutsch (1970), we present a generalization
of the rotational velocity distribution in the spirit of Tsallis
and Kaniadakis statistics. A brief discussion on the stellar
sample is made in Sec. III. Our main results are discussed in Sec.
IV and we summarize the main conclusions in Sec.V.

\section{Tsallis and Kaniadakis distribution functions}

During the last two decades a great interest has arisen in the
non-extensive Tsallis statistical mechanics (Gell-Mann \& Tsallis
2004) and, more recently, on the Kaniadakis extensive
generalized power-law statistics (Kaniadakis 2002; 2005) motivated
by various restrictions to the applicability of classical,
extensive statistical mechanics.

From the mathematical point of view, Tsallis statistics is based
on the $q$-exponential and $q$-logarithm functions, which are
defined by
\begin{equation}\label{expq}
\exp_{q}(f)= (1 + (1-q)f)^{1/{1-q}},
\end{equation}
\begin{equation}\label{expq1}
\ln_{q}(f)= {f^{1-q}-1\over 1-q},
\end{equation}
whereas the $q$-entropy associated with the $q$-statistics is
given by (Gell-Mann \& Tsallis 2004)
\begin{equation}\label{firstq}
S_q = - \int d^3 p f\ln_q f =  - \langle{\ln_q (f)\rangle}.
\end{equation}
The expressions above reduces to the standard results in the limit
$q=1$.

Similarly, the $\kappa$-framework is based on $\kappa$-exponential
and $\kappa$-logarithm functions, defined as
\begin{equation}\label{expk}
\exp_{\kappa}(f)= (\sqrt{1+{\kappa}^2f^2} + {\kappa}f)^{1/{\kappa}},
\end{equation}
\begin{equation}\label{expk1}
\ln_{\kappa}(f)= {{f^{\kappa}-f^{-\kappa}\over 2\kappa}}.
\end{equation}
The $\kappa$-entropy is given by (Kaniadakis 2002; 2005)
\begin{equation}\label{first}
S_\kappa = - \int d^3 p f\ln_\kappa f =  - \langle{\ln_\kappa
(f)\rangle}.
\end{equation}
Again, the standard results are attained in the limit $\kappa=0$.

The Tsallis statistics has been investigated in a wide range of
problems in physics \footnote{For a complete and updated list of
refences see http://tsallis.cat.cbpf.br/biblio.htm}. In the
astrophysical domain, the first applications of this powerlaw
statistics studied stellar polytropes (Plastino \& Plastino 1993)
and the peculiar velocity function of galaxy clusters (Lavagno
1998). More recently, Kaniadakis statistics (Kaniadakis 2002;
2005) has also been studied in the theoretical and experimental
context, however the first application with a possible connection
with astrophysical system has been the simulation in relativistic
plasmas (Lapenta et al. 2008). More recently, by considering the
distribution of stellar rotational velocity for low-mass stars in
the Pleiades open cluster, Carvalho et al. (2008) showed that
$\kappa$ and $q$ distributions gives a good fit for the observed
distribution.

Following the model given by the authors (Soares et al. 2006 and
Carvalho et al. 2008), it is possible to show that the Kaniadakis
and Tsallis distributions are given by

\begin{equation}
 F(\Omega)=\exp_\kappa\left({-\Omega^2\over\sigma_{\kappa}^2}\right),
 \quad F(\Omega)=\exp_q\left({-\Omega^2\over\sigma_{q}^2}\right)
\end{equation}
where $\Omega$ is the non-dimensional quantity $\omega j$, $j$
being a parameter with the dimension $\omega^{-1}$.

Here, it is worth mentioning that the standard distribution of the
true rotational velocity $V$ for a star sample is $F(V)\sim
V^2\exp(-V^2)$. As shown by Deutsch (1970), the standard observed
distribution of the projected rotational velocity $V sin i$, for a
random orientation of axes, must be given by $\phi(y)\sim
y\exp(-y^2)$ (Kraft 1970), with  $y = V sin i$. Henceforth, the
$\kappa$-distributions $\phi_\kappa(y)$ ($q$-distributions
$\phi_q(y)$) should reproduce the standard one, in the same way as
$F_\kappa(v) (F_q(v))$ recovers $F(V)$ in the $\kappa=0$ $(q=1)$
limiting case. Therefore, by considering these arguments, we
introduce the following distribution function for the observed
stellar rotational velocities

\begin{equation}
\label{kappadistY}
 \phi_\kappa(y)= y \exp_\kappa \left(-{y^2\over\sigma_{\kappa}^2}\right),
 \quad \phi_q (y)=y \exp_q\left(-{y^2\over\sigma_{q}^2}\right).
\end{equation}

\section{The stellar sample}
\label{sample}

The rotational velocities $V\sin{i}$ used in the present analysis
were taken from the rotational all-sky, magnitude-limited and
kinematically unbiased survey of more than 16,000 nearby F and G
dwarf stars from the Nordstr$\ddot{o}$m et al. (2004).
Nevertheless, it is important to underline that all the
observational data used in this study were taken from Holmberg et
al. (2007), which brings an improved version of the previous
catalogue, with new calibrations for all the relevant stellar
parameters. All the selected objects are low-mass stars in the
$b-y$ range ($0.2 - 0.6$) corresponding to an effective
temperature ranging
 from 4500 K to 7800 K  and a mass range $0.65 M_{\bigodot} - \sim
2.0 M_{\bigodot}$. The low-mass limit at $0.65 M_{\bigodot}$
reflects the red colour cutoff. Individual error estimation for
all masses in the catalogue is about $0.05 M_{\bigodot}$.  For a
complete discussion on the observational procedure, calibration
and error analysis the reader is referred to Nordstr$\ddot{o}$m et
al. (2004).

The catalogue comprises a sample of 16,682 stars containing visual
double stars and  spectroscopic binaries of all kinds. The total
number of binary stars of any type is 5,622. The sample thus
comprises 11,060 stars that are not known to be double; of these,
7,817 have measured radial velocities consistent with their being
true single stars.

The projected rotational velocities $V\sin{i}$ are only given to
the nearest $km/s$, and from 30 $km/s$ and upwards only to the nearest
5 or 10 $km/s$. Sense the great majority of stars have rotations
below 20 $km/s$, for the present analysis we have thus selected
11,818 stars with $V\sin{i}< 30$ $km/s$ which we shall call the "original sample".

The completness limit of the sample is a weak function of $b-y$.
The distribution of the photometric $V$ magnitude for all stars
shows that the original sample begins to depart from completeness
near $V=7.8$. This is the value we adopt in this work. A total of
4,473 stars have magnitude $V\leq 7.8$, satisfy the completness
criterium, and we define these as our "complete sample".

\section{Results}

Both Tsallis and Kaniadakis generalized maxwellian are two
parameter ($q - \sigma_q$ and $\kappa - \sigma_\kappa$) nonlinear
functions given by (\ref{kappadistY}). Since our aim is to test
the deviation of the velocity distribution from the standard
maxwellian we have fitted a $q$-maxwellian (Tsallis) and a
$\kappa$-maxwellian (Kaniadakis) distribution to the observed
distribution using a nonlinear least squares fitting. We apply a
nonlinear regression method based on the Levenberg-Marquardt
algorithm. The value of the distribution parameters together with
the reduced $\chi^2$ are given in Table 1.

We can clearly see that the $V\sin{i}$ distribution of main
sequence field stars in this sample does not obey a standard
Maxwellian function since the values of  $q$ and $\kappa$ are
significantly different from 1 and 0, respectively.

\begin{figure}[th]
\vspace{.05in}\centerline{\psfig{figure=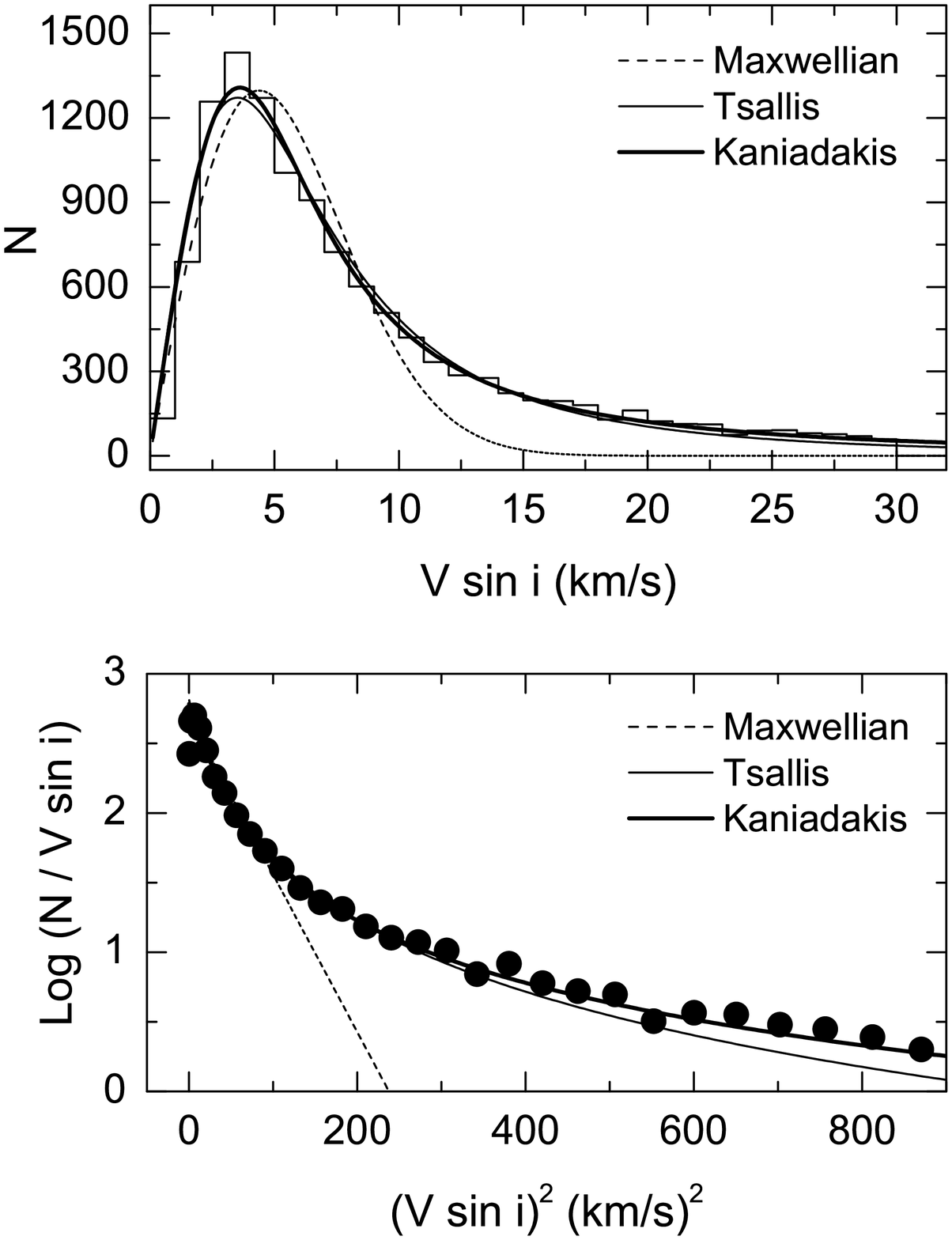,width=4.0truein,height=5.5truein}\hskip
0.05in}
\caption{Observed
distribution of the rotational velocity for all F and G field
stars in the original sample. In the upper panel we show the histogram
(number of stars $N$ in each $1~km/s$ bin as a function of
$V\sin{i}$), whereas in the lower panel, we show the logarithm
 of $N / V\sin{i}$ as a function of $(V\sin{i})^2$.
 The curves represent the best fitted maxwellian (dashed
line), Tsallis (thin line) and Kaniadakis (thick line)
distribution. The fitting parameters are in Table 1.} \label{Fig1}
\end{figure}

Interestingly, the values of  $q$ and $\kappa$ do not change
whether we consider the original sample or the complete sample.
Nevertheless, the widths of the distributions are notably
different. For the original sample $\sigma_q=4.25\pm 0.21$ and
$\sigma_\kappa=4.95\pm 0.15$, while for the complete sample we
have $\sigma_q=5.08\pm 0.25$ and $\sigma_\kappa=5.92\pm 0.19$.
Even though the difference is significant, it is within the
precision of the projected rotational velocities measurement, that
is, $1~km/s$.

In the upper panel of Fig. \ref{Fig1} we show the best fits for
the histogram of the observed distribution of $V\sin{i}$ according
the results in Table 1 for the original sample. The $q$ and
$\kappa$-Maxwellian functions are represented, respectively by the
thin and thick lines. The dashed line represents the standard
Maxwellian function. The distribution of observed $V\sin{i}$ is
without a doubt more adequately fitted by either a $q$ and
$\kappa$-Maxwellian function. This can be more clearly seen in the
lower panel of Fig. \ref{Fig1} where we have plotted the logarithm
of the distribution divided by $V\sin{i}$, that is
$log(\phi(y)/y)$, as a function of $(V\sin{i})^2$ so that the
standard Maxwellian is represented by a straight (dashed) line.
The Kaniadakis distribution is represented by the thick line while
Tsallis distribution by the thin line. We observe that both
non-gaussian distributions fit well the observed data, although
the Kaniadakis function fits the data slightly better than Tsallis
function. This can be seen in Table 1 where the values of the
reduced-$\chi^2$ is always smaller for the Kaniadakis fit compared
with the Tsallis one. Also, the uncertainties in the parameters of
the Kaniadakis distribution are smaller than in the case of
Tsallis distribution.

\begin{figure}[th]
\vspace{.05in}\centerline{\psfig{figure=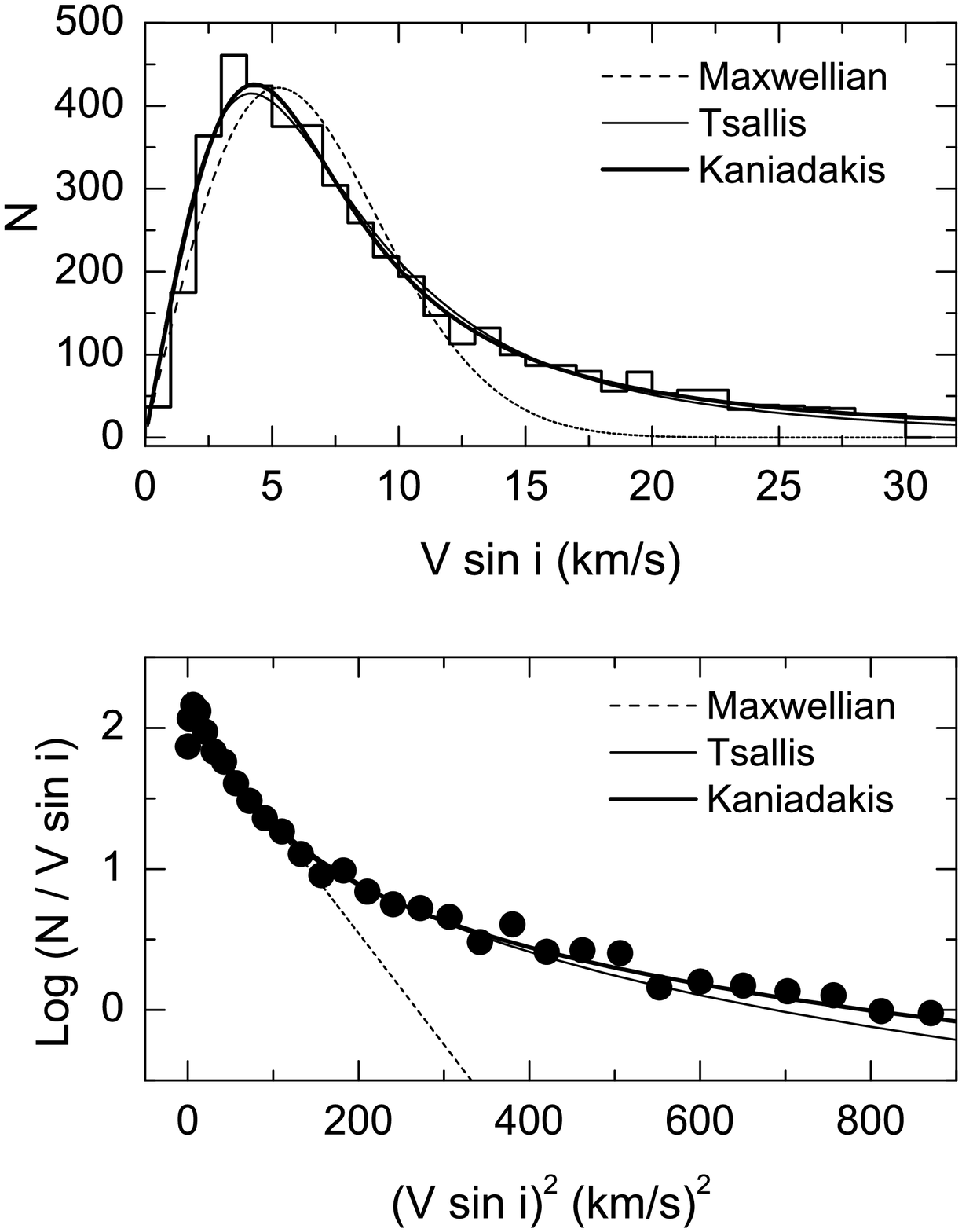,width=4.0truein,height=5.5truein}\hskip
0.05in}
\caption{Observed
distribution (histogram) of the rotational velocity for all F and
G field stars in the complete sample. In the upper panel we show
the histogram (number of stars $N$ in each $1~km/s$ bin as a
function of $V\sin{i}$), whereas in the lower panel, we show the
logarithm
 of $N / V\sin{i}$ as a function of $(V\sin{i})^2$.
 The curves represent the best fitted maxwellian (dashed
line), Tsallis (thin line) and Kaniadakis (thick line)
distribution. The fitting parameters are in Table 1.} \label{Fig2}
\end{figure}

In  Fig. \ref{Fig2} we present the fitting to the complete sample
using the parameter given in Table 1. The results are similar to
the original sample. The $q$ and $\kappa$-distribution give the best
fit with Kaniadakis function being slightly better.

Finally, we have found no noteworthy difference on the
distribution of the projected rotational velocity between single
and binary stars (see Table 1). Despite the fact that the values
of $q$ and $\sigma_q$ and $\kappa$ and $\sigma_\kappa$ are
slightly larger for binary stars, the difference are, in most
case, within the statistical fluctuation.

\section{Conclusions}

In this work we have used non-gaussian statistics to investigate
the observed distribution of projected rotational velocity of a
magnitude complete sample of more than 16,000 nearby F and G dwarf
stars. We have fitted a standard maxwellian and the $q$ and
$\kappa$-maxwellian based on the generalization of the power-law
statistics proposed by Tsallis and Kaniadakis. We conclude that
the $V\sin{i}$ distribution deviates significantly from a standard
maxwellian and the best fit is attained for Tsallis distribution
with $q=1.521\pm 0.043$ and Kaniadakis distribution with
$\kappa=0.667\pm 0.033$. These values correspond to all stars
(single + binary) in the complete sample.

As far as we are aware, this is the first time that Tsallis and
Kaniadakis statistics are tested for such an exceptionally large
sample of stars. This gives us an excellent degree of confidence
on the fact that the distribution of stellar rotational velocity
does not follow a Maxwellian-Boltzmann law as suggested by the
pioneer studies by Chandrasekhar and Munch (1950) and Deutsch
(1970). The present results show clearly that for the stellar
rotation the non-extensivity holds, and that the distribution of
the observed rotational velocity is explained by a generalized
power--law statistics in the spirit of Tsallis and Kaniadakis
statistical mechanics.

% \vspace{0.5cm}

\noindent {\bf Acknowledgments:} The authors are partially supported by the CNPq and
FAPERN Brazilian Agencies.

%%%%%%%%%%%%%%%%%%%%%%%%%%%%%%%%%%%%%%%%%%%%%%%%%%%%%%%%%%%%%%%%%%%%%%%%%%
%%%%%%%%%%%%%%%%%%%%%%%%%%%%%     TABLE I     %%%%%%%%%%%%%%%%%%%%%%%%%%%%
%%%%%%%%%%%%%%%%%%%%%%%%%%%%%%%%%%%%%%%%%%%%%%%%%%%%%%%%%%%%%%%%%%%%%%%%%%
\begin{table}[tbp]
\tablecolumns{1}
\tiny
\tablewidth{0pt}
\begin{center}
\caption[]{Best values of the parameters of Tsallis ($q$ and
$\sigma_q$) and Kaniadakis ($\kappa$ and $\sigma_{\kappa}$)
distribution determined using $\chi^2$ test for the rotational
velocity of stars in the Nordstr$\ddot{o}$m et al. catalogue.} \label{tab1}
\vspace{0.4in}
\begin{tabular}{cc|ccc||ccc}
\hline\hline
        &  N   & $q$   &  $\sigma_q$ (km/s)  & reduced-$\chi^2$ &  $\kappa$   &
$\sigma_{\kappa}$ (km/s) & reduced-$\chi^2$ \\
 \hline
 \hline
&&&&&&&\\ {\bf Original Sample}  &&&&&&&\\

   all stars   & 11818 & $1.525\pm 0.041$ & $4.25\pm 0.21$ & 4493 & $ 0.667\pm
0.029$ & $4.95\pm 0.15$ & 3088 \\
     single    & 6888  & $1.484\pm 0.053$ & $4.13\pm 0.25$ & 422  & $ 0.636\pm
0.038$ & $4.74\pm 0.17$ & 400 \\
     binary    & 4930  & $1.551\pm 0.037$ & $4.63\pm 0.21$ & 223  & $ 0.687\pm
0.031$ & $5.47\pm 0.17$ & 207 \\
 &&&&&&&\\
 \hline
 &&&&&&&\\
{\bf Complete Sample} &&&&&&&\\
   all stars   & 4473 & $1.521\pm 0.043$ & $5.08\pm 0.25$ & 550 & $0.667\pm 0.033$ &
$5.92\pm 0.19$ & 424 \\
     single    & 2382 & $1.507\pm 0.046$ & $4.92\pm 0.26$ & 196 & $0.655\pm 0.035$ &
$5.69\pm 0.19$ & 142 \\
     binary    & 2091 & $1.529\pm 0.049$ & $5.34\pm 0.31$ & 149 & $0.672\pm 0.041$ &
$6.26\pm 0.25$ & 133 \\
 &&&&&&&\\
\hline \hline
\end{tabular}
\end{center}
\end{table}


\begin{thebibliography}{}

%\bibitem[Alexander(2006)]{2006Sci.320..1617} Alexander, C.~M.~O., Grossman, J.~N., Ebel, D.~S., \& Ciesla, F.~J. \ 2006, Science, 320, 1617

\bibitem{carvalho2008}Carvalho, J. C., Silva, R., do Nascimento, J. D.Jr., De Medeiros, J. R.
 2008,  Europhysics Letters,  84, 59001


\bibitem{munch50} Chandrasekhar, S.,  Munch, G.\ 1950,  \apj,   111, 142

\bibitem{medeiros96} De Medeiros, J. R., Da Rocha, C., Mayor, M. 1996, \aap,  314, 499

\bibitem{deut70} Deutsch, A.,~J. 1970, in: A Slettebak (Ed.), Stellar Rotation, IAU
Colloquium, Reidel, Dordrecht,  p. 207.

\bibitem{dwo74} Dworetssky, M., M. 1974, \apj,  28, 101

\bibitem{tsallis1} Gell-Mann, M., and Tsallis C., (Eds.), Nonextensive Entropy -
Interdisciplinary Applications, Oxford University Press, New York,
(2004).

\bibitem{holmberg07}Holmberg, J., Nordstr$\ddot{o}$m, B.,  Andersen, J. 2007, \aap, 475, 519

\bibitem{k1} Kaniadakis,~G.\ 2002, Phys. Rev. E,  66, 056125

\bibitem{k2} Kaniadakis,~ G.\ 2005, Phys. Rev. E,  72, 036108

\bibitem{raft70} Kraft, R. P.  1970, in {\it Spectroscopic
Astrophysics. An Assessment of the Contributions
of Otto Struve}, G. H. Herbig, ed. (Berkeley: University of
California Press),  p.385


\bibitem{lavagno98} Lavagno, A., Kaniadakis, G.,  Rego-Monteiro, M.,   Quariti,  P.,
Tsallis, C. 1998,  Astrophys. Lett. Commun., 35, 449

\bibitem{lapenta07} Lapenta, G., Markidis, S., Marocchino, A., and  Kaniadakis, G. 2007,
\apj,   666, 949


\bibitem{plastino93} Plastino, A.,  \& Plastino, A.~R. \ 1993, Phys. Lett. A,  174, 384

\bibitem{nord2004} Nordstr$\ddot{o}$m, B., Mayor, M., Andersen, J., Holmberg, J., Pont, F.,
   Jorgensen, B. R., Olsen, E. H., Udry, S., Mowlavi, N.  A. 2004, \aap, 418, 989

\bibitem{Queloz98} Queloz, D.,  Allain, S.,  Mermilliod, J. -C.,
Bouvier J., Mayor, M. 1998, \aap, 335, 183


\bibitem{soares06} Soares, B.~B., Carvalho, J.~C., do Nascimento, J.~D.~Jr. and De
Medeiros, J. 2006, Physica A,  364, 413

\bibitem{struve45} Struve, O. \ 1945,  Pop.  53, 202

\bibitem{Wolff82}  Wolff, S.~C., Edwards, S., Preston, G. W. 1982, \apj,  252,  322


\end{thebibliography}
\end{document}